\documentclass[reprint,a4paper,showpacs,showkeys,pra,aps]{revtex4-1}
\usepackage[utf8]{inputenc}
\usepackage[english]{babel}
\usepackage{graphicx}
\usepackage{amsmath}
\usepackage{amssymb}
\usepackage{textcomp}
\usepackage{subfigure}

\newcommand{\Biz}{$\textrm{Bi}^{0}$}
\newcommand{\Bipi}{$\textrm{Bi}^{+}$}
\newcommand{\Biiip}{$\textrm{Bi}_2^{+}$}
\newcommand{\BiVacI}{\mbox{$\textrm{Bi}^{0}\cdots\textrm{V}_{\textrm{I}}^{0}$}}
\newcommand{\BiVacCl}{%
\mbox{$\textrm{Bi}^{+}\cdots\textrm{V}_{\textrm{Cl}}^{-}$}}
\newcommand{\BiVacz}[1]{\mbox{$\textrm{Bi}\cdots\textrm{V}_{\textrm{#1}}$}}
\newcommand{\TlCl}{\mbox{TlCl}}
\newcommand{\TlClBi}{\mbox{TlCl:Bi}}
\newcommand{\CsI}{\mbox{CsI}}
\newcommand{\CsIBi}{\mbox{CsI:Bi}}
\newcommand{\Term}[4]{\mbox{${}^{#1}{\textrm{#2}}_{#3}^{#4}$}}
\newcommand{\Cub}[1]{\mbox{{#1}m$\overline{3}$m}}
\newcommand{\QE}{\mbox{Quantum-Espresso}}
\newcommand{\Elk}{\mbox{Elk}}
\newcommand{\cminv}{$\textrm{cm}^{-1}$}
\newcommand{\mkm}{\textmu{}m}
\begin{document}
\title{%
Centers of near-infrared luminescence \\
in bismuth-doped \mbox{TlCl}{} and \mbox{CsI}{}
crystals}
\author{V.~O.~Sokolov}
\email[E-mail:~~]{vence.s@gmail.com}
\author{V.~G.~Plotnichenko}
\author{E.~M.~Dianov}
\affiliation{Fiber~Optics~Research~Center of the~Russian~Academy~of~Sciences \\
38~Vavilov~Street, Moscow 119333, Russia}
\begin{abstract}
A comparative first-principles study of possible bismuth-related centers in
\TlCl{} and \CsI{} crystals is performed and the results of computer modeling
are compared with the experimental data of Refs.~\cite{We13, Su11, Su12}. The
calculated spectral properties of the bismuth centers suggest that the IR
luminescence observed in \TlClBi{} \cite{We13} is most likely caused by
\BiVacCl{} centers (\Bipi{} ion in thallium site and a negatively
charged chlorine vacancy in the nearest anion site). On the contrary, \Bipi{}
substitutional ions and \Biiip{} dimers are most likely responsible for the IR
luminescence observed in \CsIBi{} \cite{Su11, Su12}.

\end{abstract}
\pacs{%
22.70.-a,  
42.70.Hj,  
78.55.-m,  
}
\maketitle

\section{Introduction}
\label{sec:Intro}
New near-IR luminescence, quite different from much-studied Bi$^{3+}$-related
luminescence \cite{Blasse68, Weber73}, was discovered in 2001 in bismuth-doped
aluminosilicate glass \cite{Fujimoto01}. Then optical amplification at
1.3~\mkm{} was demonstrated \cite{Fujimoto03}. Ever since bismuth-doped glasses
and optical fibers based on such glasses attract a considerable interest due to
a broadband IR luminescence in the range of 1.0 -- 1.7~\mkm{} employed
successfully in fiber lasers and amplifiers (see e.g. the review
\cite{Dianov09}).

Although the origin of the IR luminescence is still not clear, recently a belief
has been strengthened that subvalent bismuth centers are responsible for
the luminescence \cite{Peng11}. In our opinion, monovalent bismuth centers are
of a particular interest.

Glasses as disordered systems are very complicated to study impurity centers
structure. In this regard, crystals may be of interest as model hosts containing
bismuth-related centers. In particular, crystalline halides of monovalent metals
are convenient hosts for monovalent bismuth centers. These crystals have a
simple structure (primitive, \Cub{P}, or face-centered, \Cub{F}, cubic lattice).
Bismuth can easily form monovalent substitutional centers in such lattice.
Similar subvalent thallium and lead centers in \Cub{F}{} crystals were studied
extensively (e.g. thallium in \mbox{KCl} \cite{Mollenauer83} and lead in
\mbox{MF$_2$}, M$\, = \,$Ca, Sr, Ba \cite{Fockele89}). By analogy, the models of
bismuth-related centers in oxide glasses for fiber optics were suggested
\cite{Dianov10}.

Bismuth-related IR luminescence in cubic halide crystals was studied for the
first time in \mbox{BaF$_2$} \cite{Su09}, then in \CsI{} (\Cub{P}) \cite{Su11,
Su12} and recently in \TlCl{} (\Cub{P}) \cite{We13}. Two models of the centers
in \CsIBi{} were suggested in \cite{Su11, Su12}, namely, a monovalent bismuth
substitutional center, \Bipi, and a dimer center, \Biiip, formed by two \Bipi{}
substitutional centers in the nearest cation sites with an extra electron.
In what follows we report the results of computer modeling of bismuth-related 
centers in \TlClBi{} and \CsIBi{} crystals.

\section{Modeling of bismuth-related centers in T\lowercase{l}C\lowercase{l}:Bi
and C\lowercase{s}I:Bi crystals}
\label{sec:Modeling}
To study the origin of the IR luminescence in \TlClBi{} and \CsIBi, we performed
a computer simulation of the structure and absorption spectra of three
bismuth-related centers possibly occurring in both crystals. Basing on the
assumptions made in \cite{Su11, Su12} and on analogy with
\mbox{$\textrm{Tl}^0\!\left(1\right)$} centers in alkaline halide crystals (see
e.g. \cite{Mollenauer83}), we studied the \Bipi{} substitutional center as the
main form of bismuth embedding in \TlCl{} and \CsI, the \Biiip{} dimer center,
and \BiVacz{Cl}{} and \BiVacz{I}{} complexes formed by the bismuth
substitutional center and anion vacancy in its first coordination shell.

The modeling was performed in a supercell approach. $3 \times 3 \times 3$
\TlCl{} or \CsI{} supercells (54 atoms) was chosen to model \Bipi{} and
\BiVacz{Cl}{} or \BiVacz{I} centers, and $3 \times 3 \times 4$ supercell (72
atoms) was used for \Biiip. In the central region of the supercell certain
cations were substituted by bismuth atoms and an anion vacancy was formed by a
removal of one chlorine or iodine atom. Charged centers were simulated changing
the total number of electrons in the supercell. Equilibrium configurations of
bismuth centers were found by a complete optimization of the supercell
parameters and atomic positions with the gradient method. All such calculations
were performed using \QE{} package \cite{QE} in the plane wave basis in the
generalized gradient approximation of density functional theory with ultra-soft
pseudopotentials built with PBE functional \cite{PBE}. The pseudopotential
sources were taken from the pslibrary~v.~0.3.0 pseudopotential library
\cite{pslibrary}.
\begin{figure*}
\subfigure[]{%
\includegraphics[width=8.8cm, bb=0 0 1100 1100]{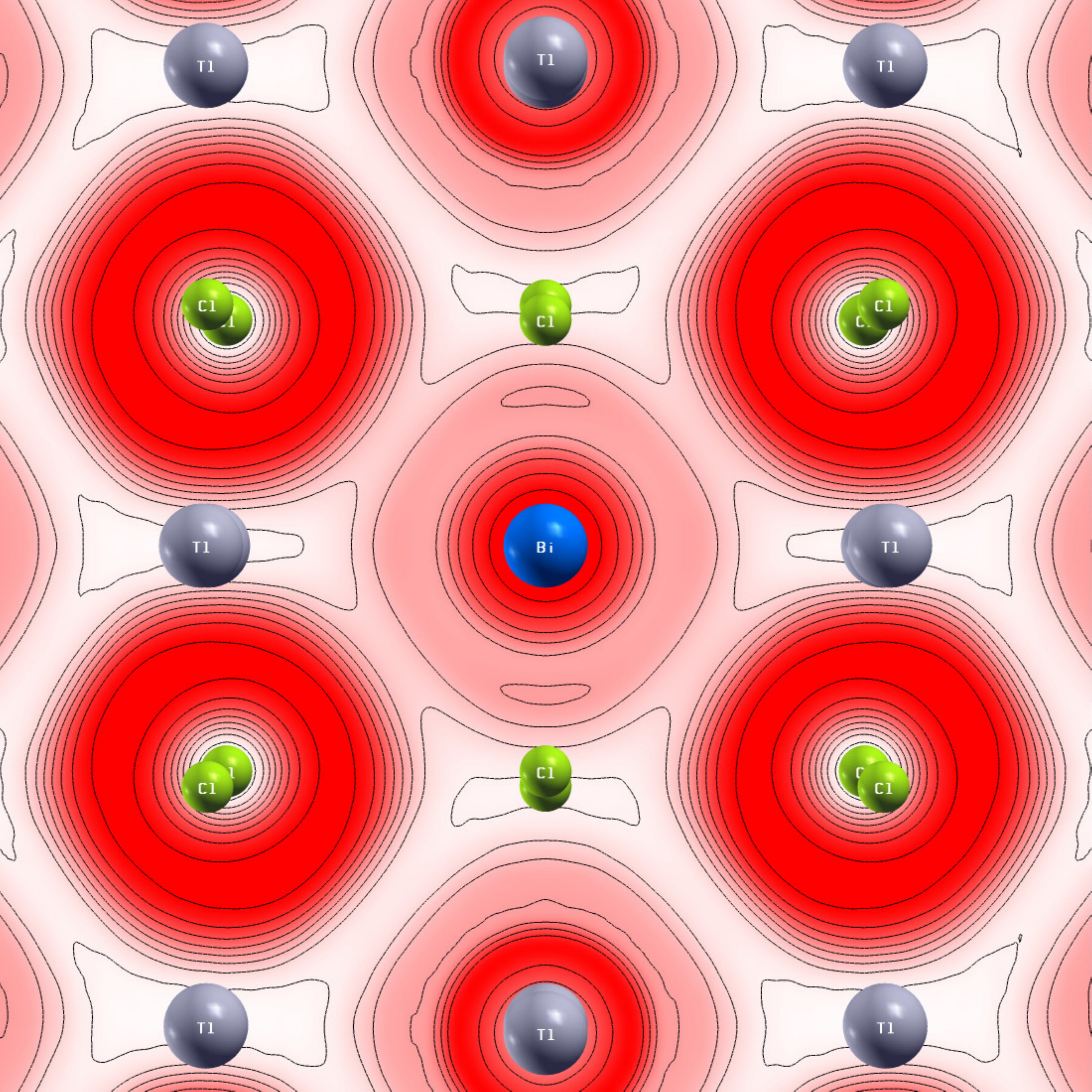}
\label{fig:Bi+_TlCl_ELF}
}
\subfigure[]{%
\includegraphics[width=8.8cm, bb=0 0 1100 1100]{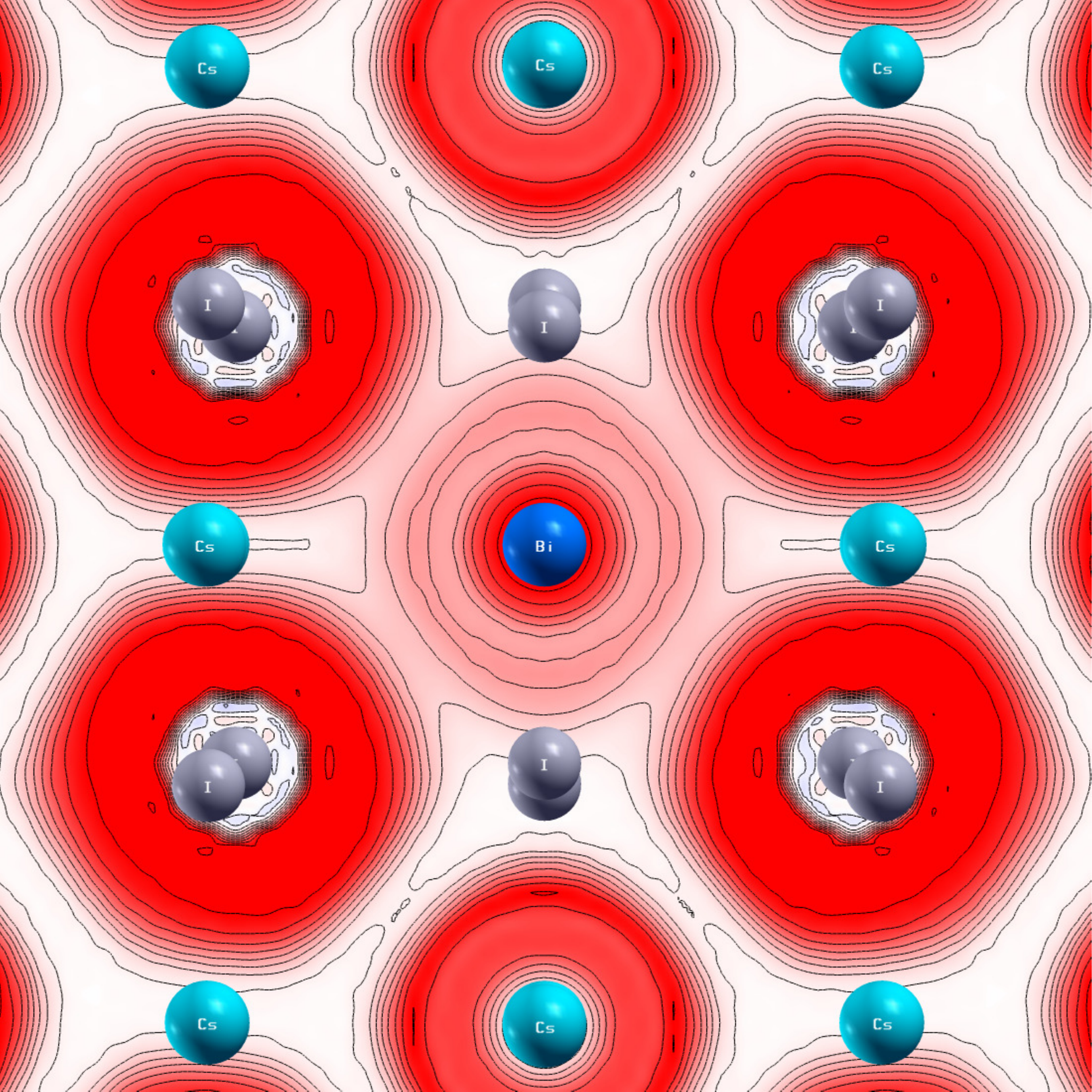}
\label{fig:Bi+_CsI_ELF}
}
\\[-0.25\baselineskip]
\subfigure[]{%
\includegraphics[width=8.8cm, bb=0 0 1100 1100]{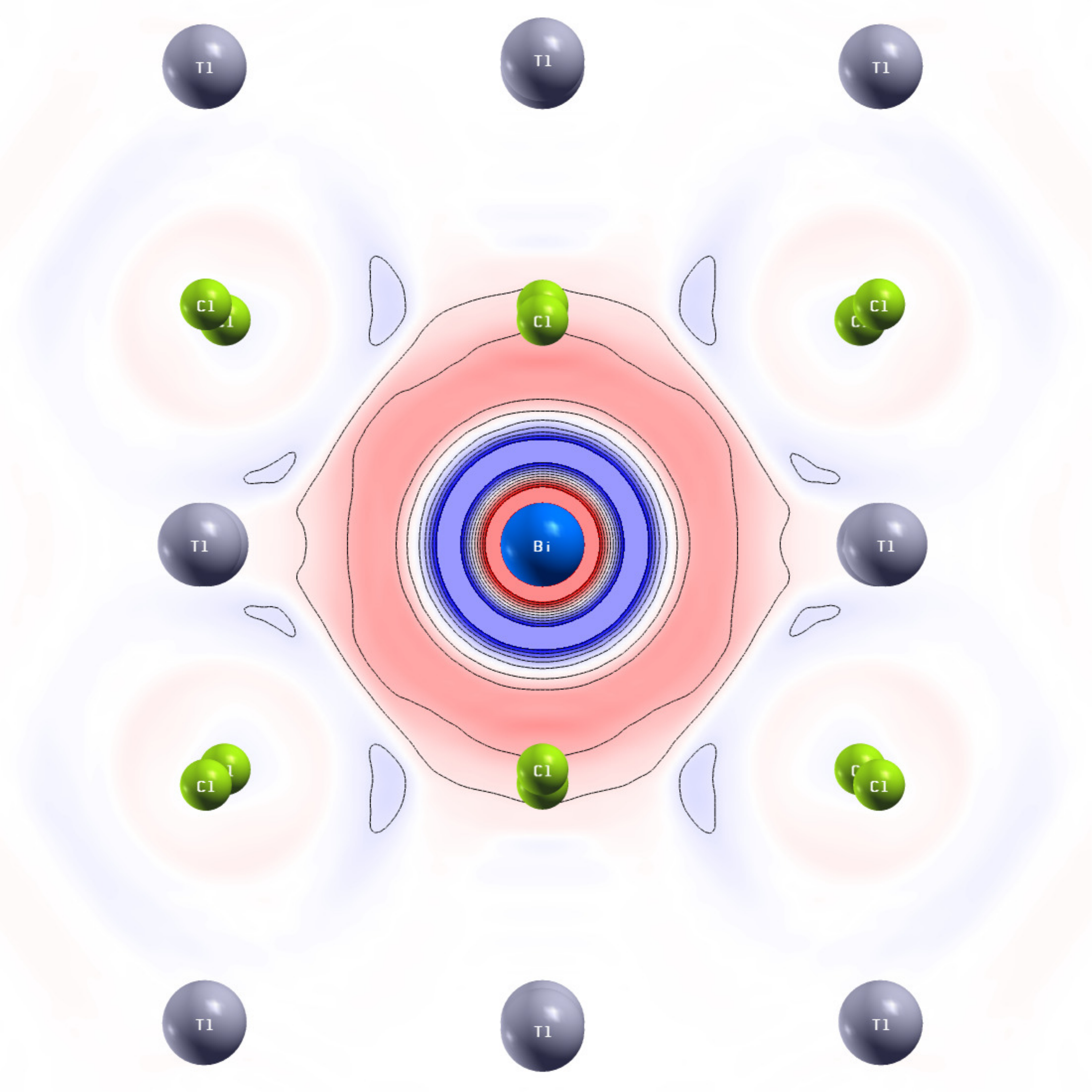}
\label{fig:Bi+_minus_TlCl_ELF}
}
\subfigure[]{%
\includegraphics[width=8.8cm, bb=0 0 1100 1100]{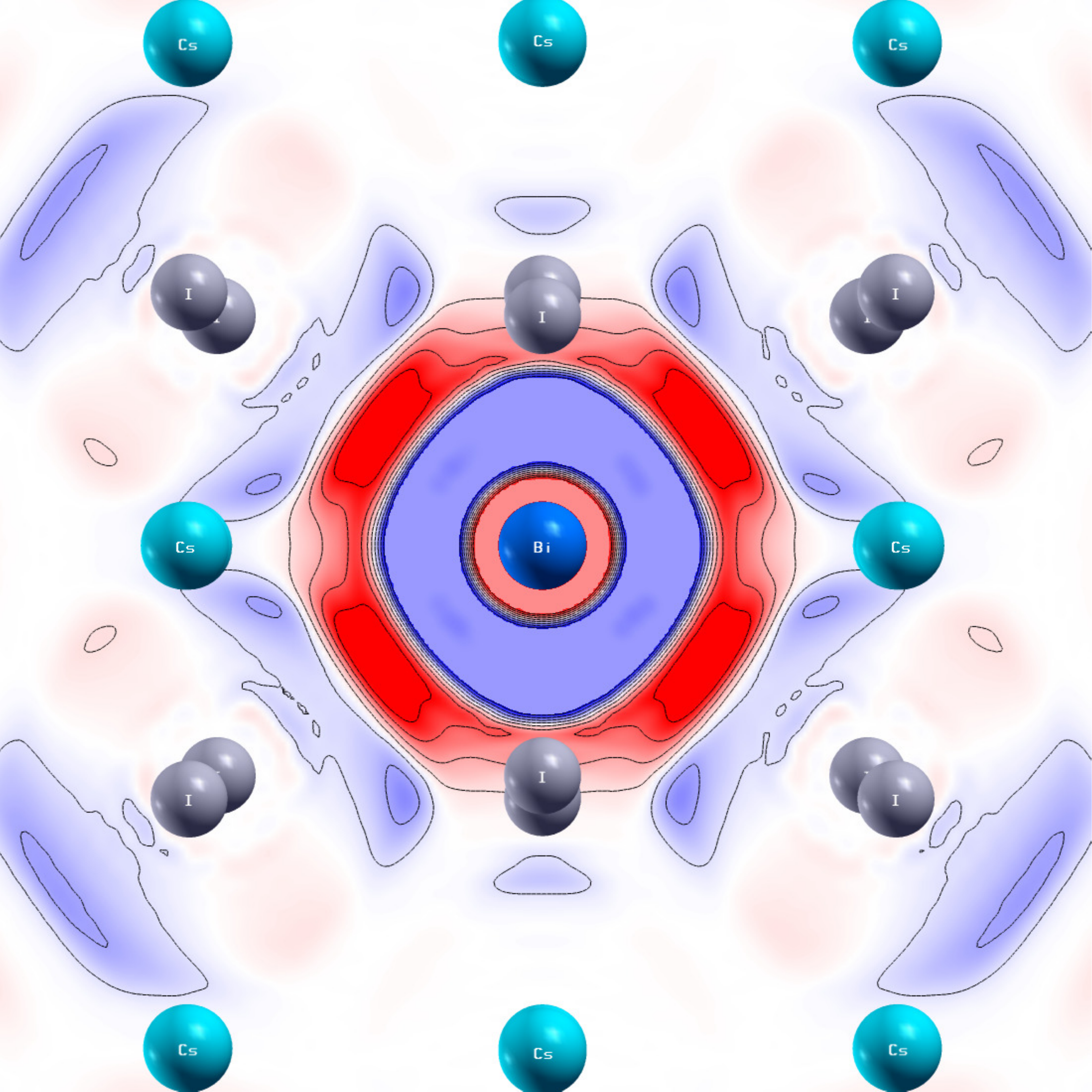}
\label{fig:Bi+_minus_CsI_ELF}
}
\caption{%
Calculated electron localization functions of \Bipi{} substitutional centers
in \subref{fig:Bi+_TlCl_ELF}~\TlCl{} and \subref{fig:Bi+_CsI_ELF}~\CsI, and
difference of calculated electron localization functions of \Bipi{}
center from those of perfect lattice: \subref{fig:Bi+_TlCl_ELF}~\TlClBi,
\subref{fig:Bi+_CsI_ELF}~\CsIBi{} (in the $\left(111\right)$ plane).
}
\label{fig:Bi+_ELFs}
\end{figure*}

To test the approach, \TlCl{} and \CsI{} lattice parameters were calculated for
the unit cell and supercells with both atomic positions and cell parameters
completely optimized. The results convergence was tested with respect to the
plane wave cutoff energy and to the $k$ points grid. The energy cutoff $\gtrsim
700$~eV and the number of $k$ points $\geq 27$ in the irreducible part of the
unit cell Brillouin zone were found to be enough to converge the total energy
within $10^{-3}$~eV per atom and to reproduce the experimental lattice
parameters with a relative accuracy of $\lesssim 1$\%. The geometry of each
supercell was reproduced with a relative accuracy better than 2\%{} with only
$\Gamma$ point of the supercell taken into account and better than 1\%{} using 8
$k$ points in the supercell in the irreducible part of the supercell Brillouin
zone. The total energy convergence was not worse than that in the case of the
unit cell.
\begin{figure*}
\subfigure[]{%
\includegraphics[width=2.79cm, bb=130 -10 285 585]{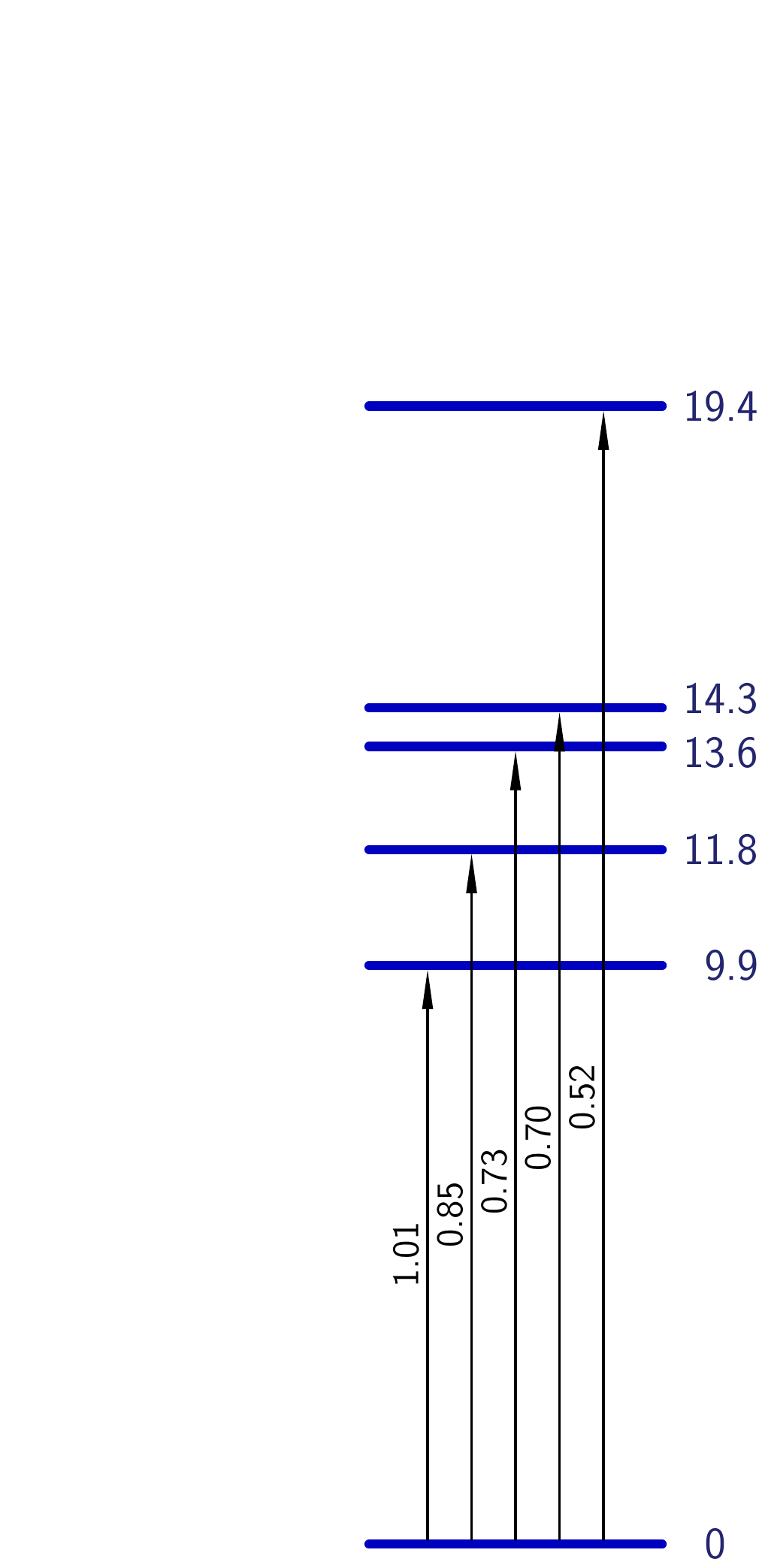}
\label{fig:Bi+_TlCl_levels}
}
\subfigure[]{%
\includegraphics[width=2.79cm, bb=130 -10 285 585]{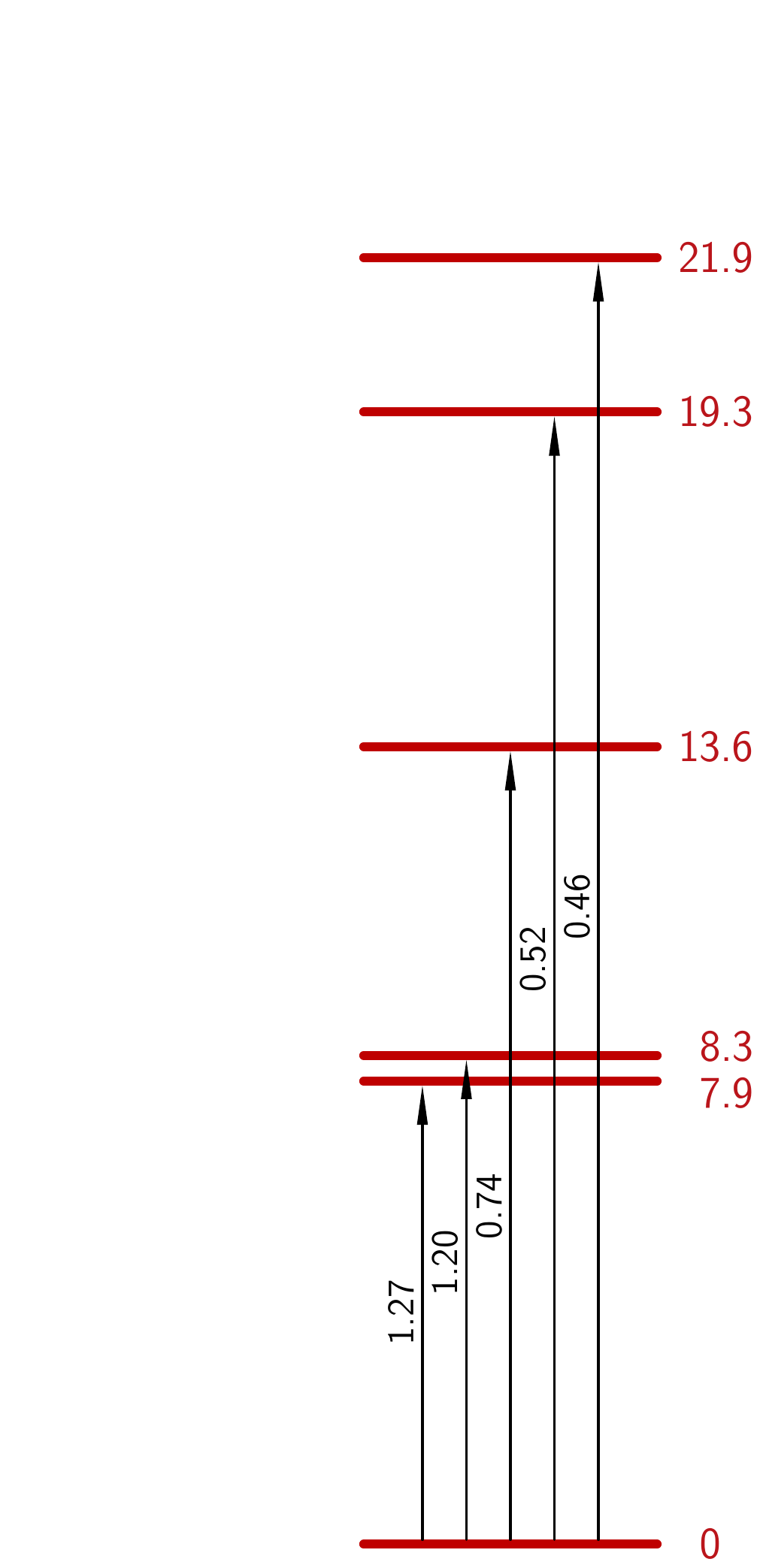}
\label{fig:Bi+_CsI_levels}
}
\subfigure[]{%
\includegraphics[width=2.79cm, bb=130 -10 285 585]{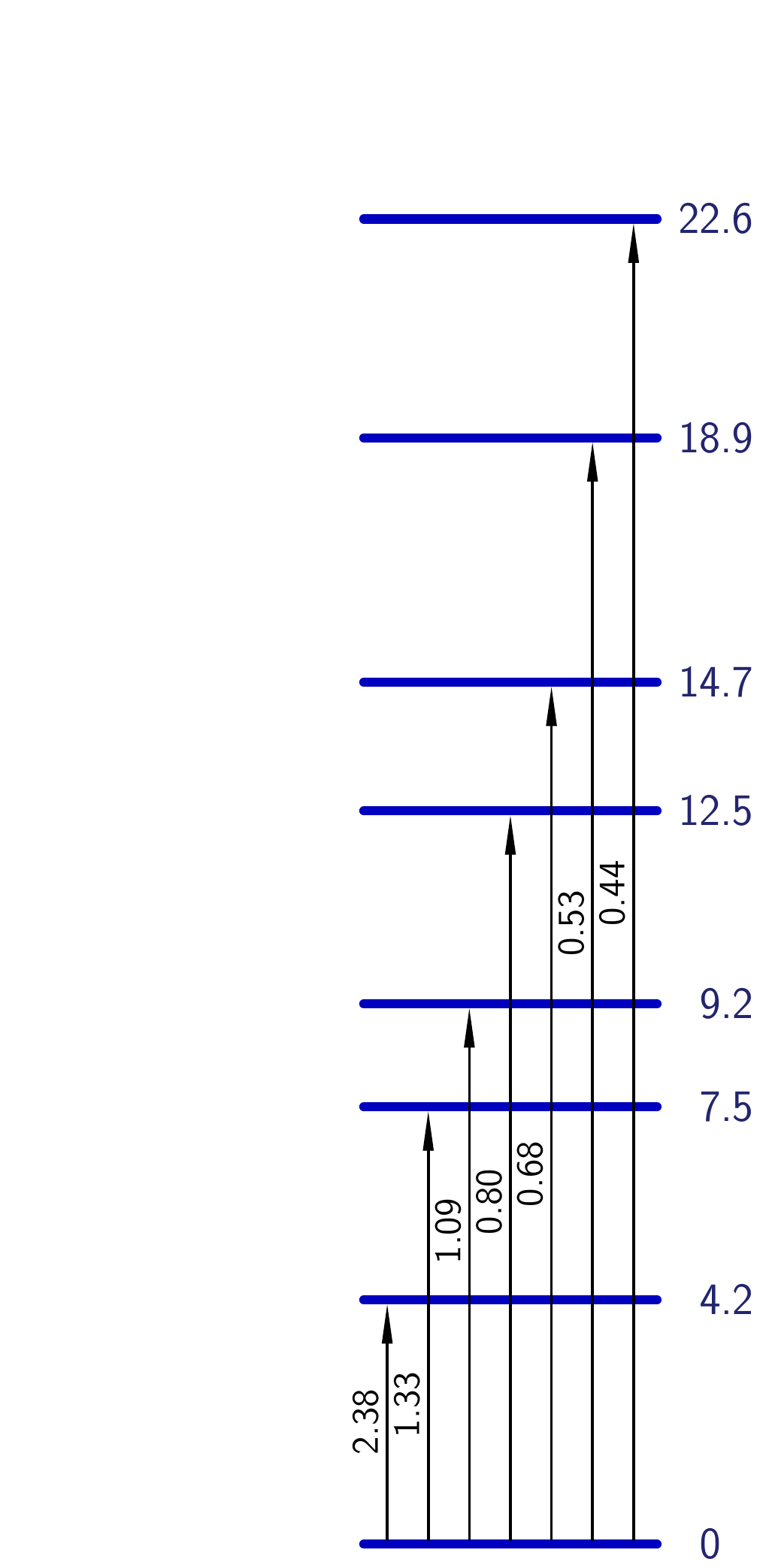}
\label{fig:Bi_2+_TlCl_levels}
}
\subfigure[]{%
\includegraphics[width=2.79cm, bb=130 -10 285 585]{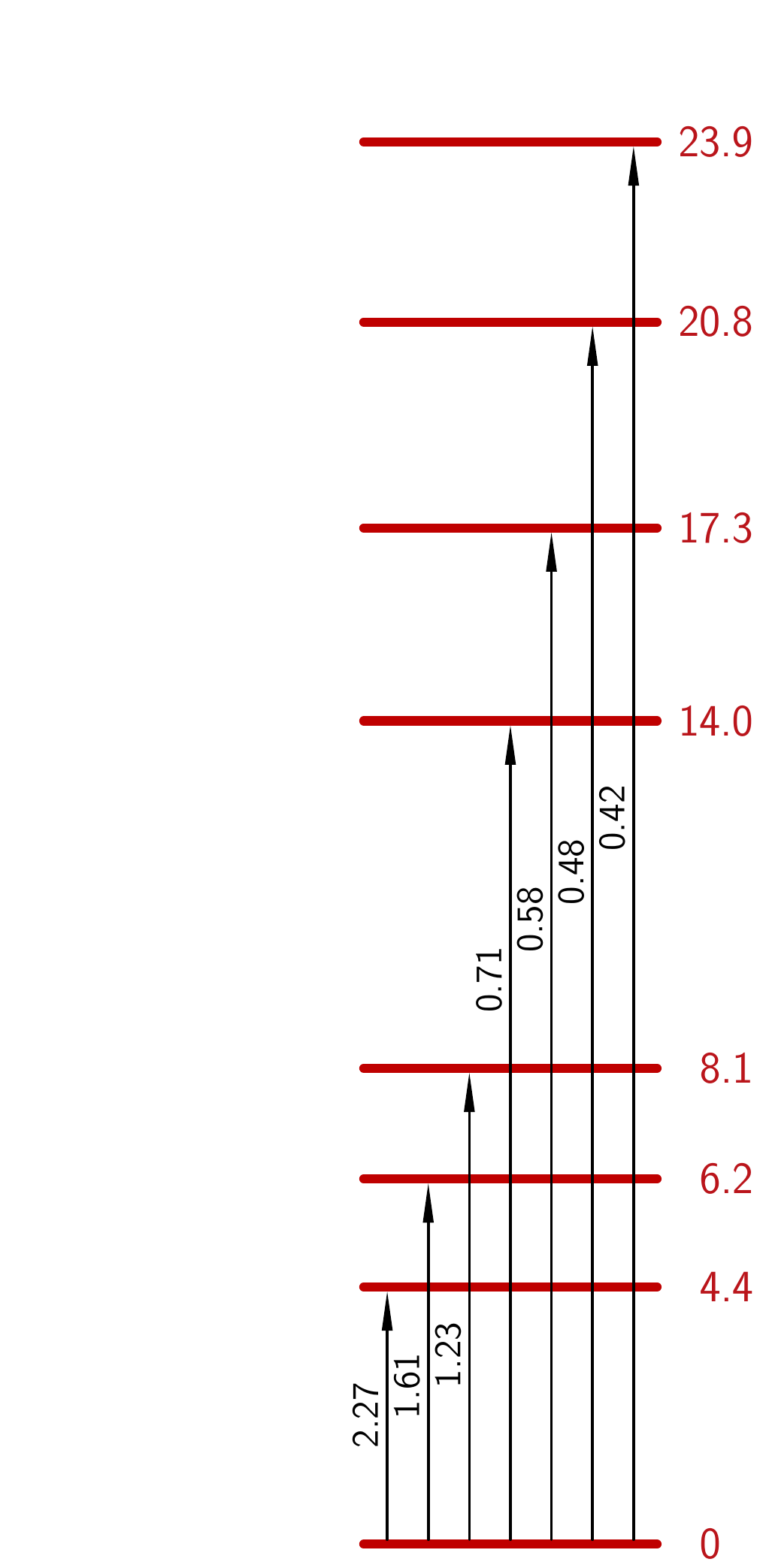}
\label{fig:Bi_2+_CsI_levels}
}
\subfigure[]{%
\includegraphics[width=2.79cm, bb=130 -10 285 585]
{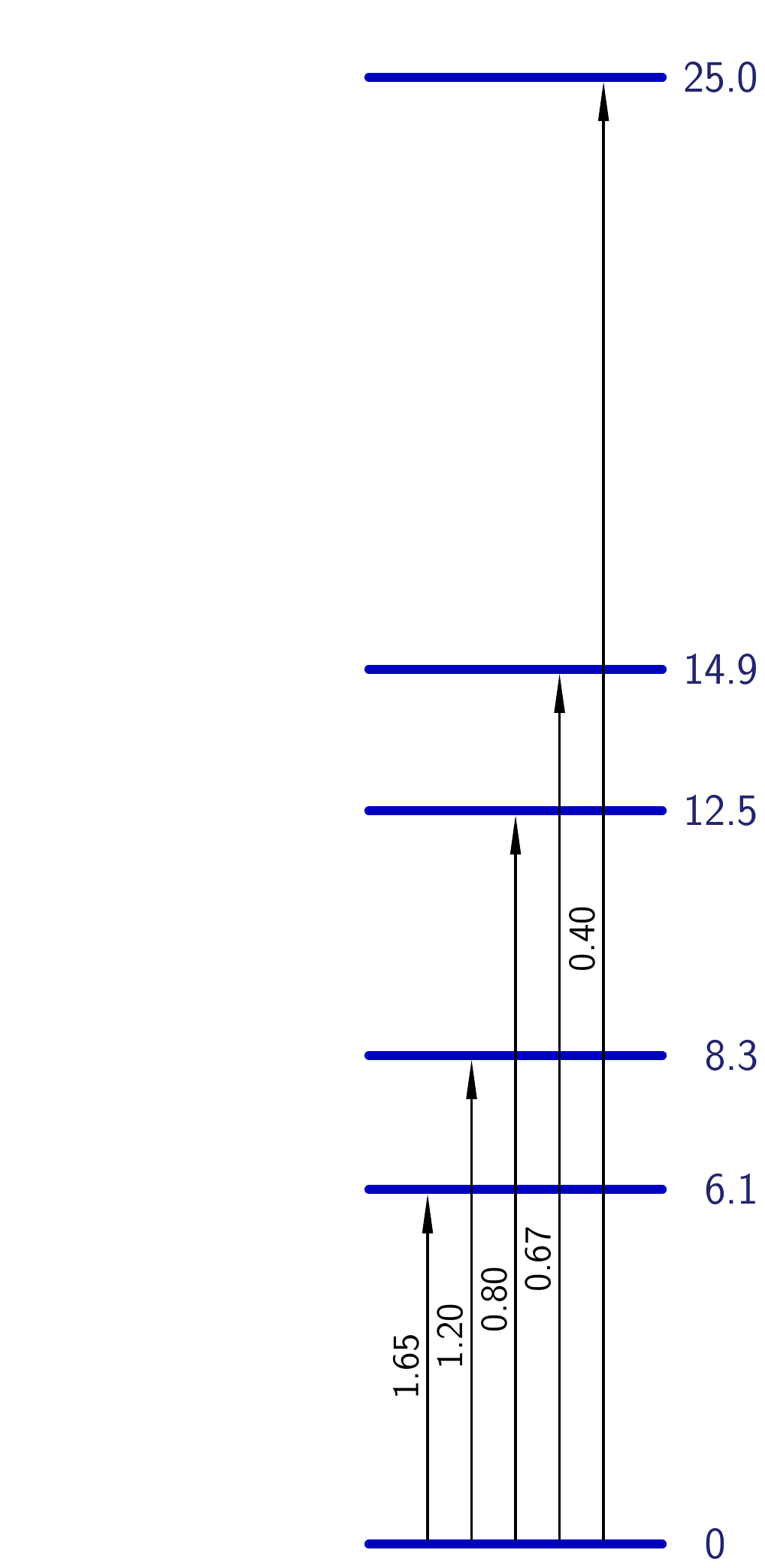}
\label{fig:Bi_and_vac_TlCl_levels}
}
\subfigure[]{%
\includegraphics[width=2.79cm, bb=130 -10 285 585]{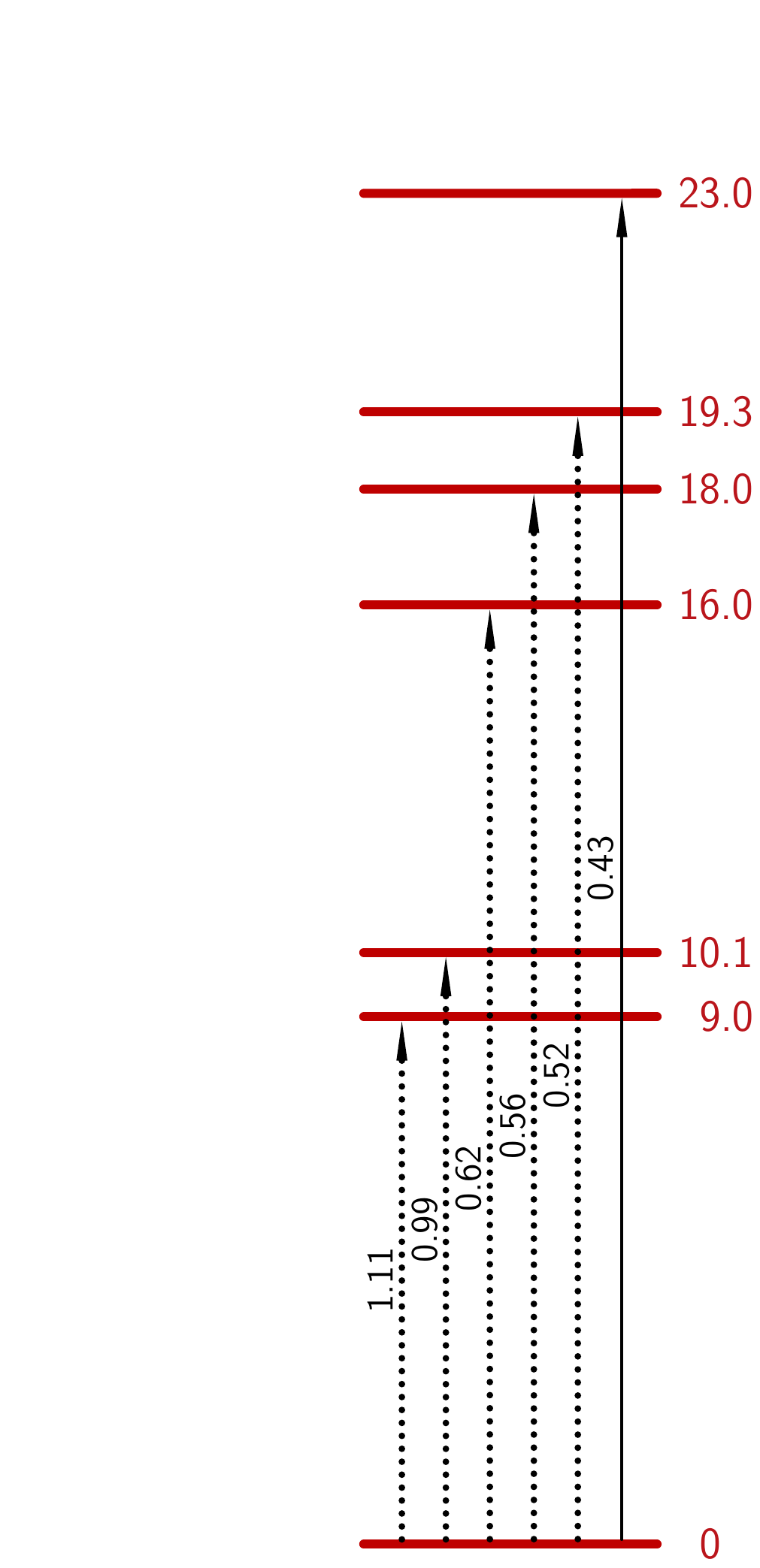}
\label{fig:Bi_and_vac_CsI_levels}
}
\caption{%
Calculated level and transition schemes of bismuth-related centers:
\subref{fig:Bi+_TlCl_levels}~\Bipi{} substitutional center in \TlClBi,
\subref{fig:Bi+_CsI_levels}~\Bipi{} substitutional center in \CsIBi,
\subref{fig:Bi_2+_TlCl_levels}~\Biiip{} dimer center in \TlClBi,
\subref{fig:Bi_2+_CsI_levels}~\Biiip{} dimer center in \CsIBi,
\subref{fig:Bi_and_vac_TlCl_levels}~\BiVacCl{} center in \TlClBi,
\subref{fig:Bi_and_vac_CsI_levels}~\BiVacI{} center in \CsIBi.
Level energies are given in $10^3$~\cminv, transition wavelengths in \mkm.
}
\label{fig:Bi_centers_levels}
\end{figure*}

Configurations of bismuth-related centers obtained by this means were used to
calculate the electron localization functions using the programs from \QE{}
package, the electron density distribution and effective charges of atoms by
Bader's method using bader~v.\,0.28 code \cite{bader}, and the absorption
spectra of the centers by Bethe-Salpeter equation method based on all-electron
full-potential linearized augmented-plane wave approach \cite{LAPW}. The
absorption spectra calculations were performed using \Elk{} code \cite{Elk} in
the local spin density approximation with PW-CA functional \cite{LSDA-PW,
LSDA-CA}. Spin-orbit interaction essential for bismuth-containing systems was
taken into account. Scissor correction was applied to transition energies
calculation with the scissor value found using modified Becke-Johnson
exchange-correlation potential known to yield accurate band gaps in many solids
\cite{Becke06, Tran09, Tran11}. The non-overlapping muffin-tin (MT) spheres of
maximal possible radii $R^\textrm{MT}$ were used. Convergence of the results was
tested with respect to plane-wave cutoff energy, to the angular momentum cutoff
for the MT density and potential, and to the $k$ points grid choice. The
plane-wave cutoff, $k_{max}$, was determined by the $R^\textrm{MT}_{min} \cdot
k_{max} = 7$ relation with $R^\textrm{MT}_{min}$ being the smallest MT radius.
The angular momentum cut-off was taken to be $l = 10$. The self-consistent
calculations were performed on the $3\times 3\times 3$ grid of $k$ points
uniformly distributed in the irreducible part of the supercell Brillouin zone.
Further increasing the cutoff and $k$ points density did not lead to significant
changes in the results. The total energy self-consistence tolerance was taken to
be $10^{-3}$~eV per atom. More dense $4\times 4\times 4$ $k$ points grid was
applied to calculate dipole matrix elements in optical spectra calculations.

Simplified configurational coordinate diagrams of bismuth-related centers were
calculated in a model restricted to the lowest excited states with a
displacement of bismuth atom(s) along $\left[111\right]$ axis for \Bipi{} and
\BiVacI{} centers and along $\left[001\right]$ axis for \Biiip{} center. In
spite of the fact that the model is inherently rough, it shows that in all the
centers studied the Stokes shift corresponding to a transition from the first
excited state to the ground one do not exceed the accuracy of the excited state
energy calculation. Hence it seems reasonable enough to estimate the IR
luminescence wavelengths by taking this Stokes shift to be zero.
\begin{figure*}
\subfigure[]{%
\includegraphics[width=8.8cm, bb=0 0 1100 1100]{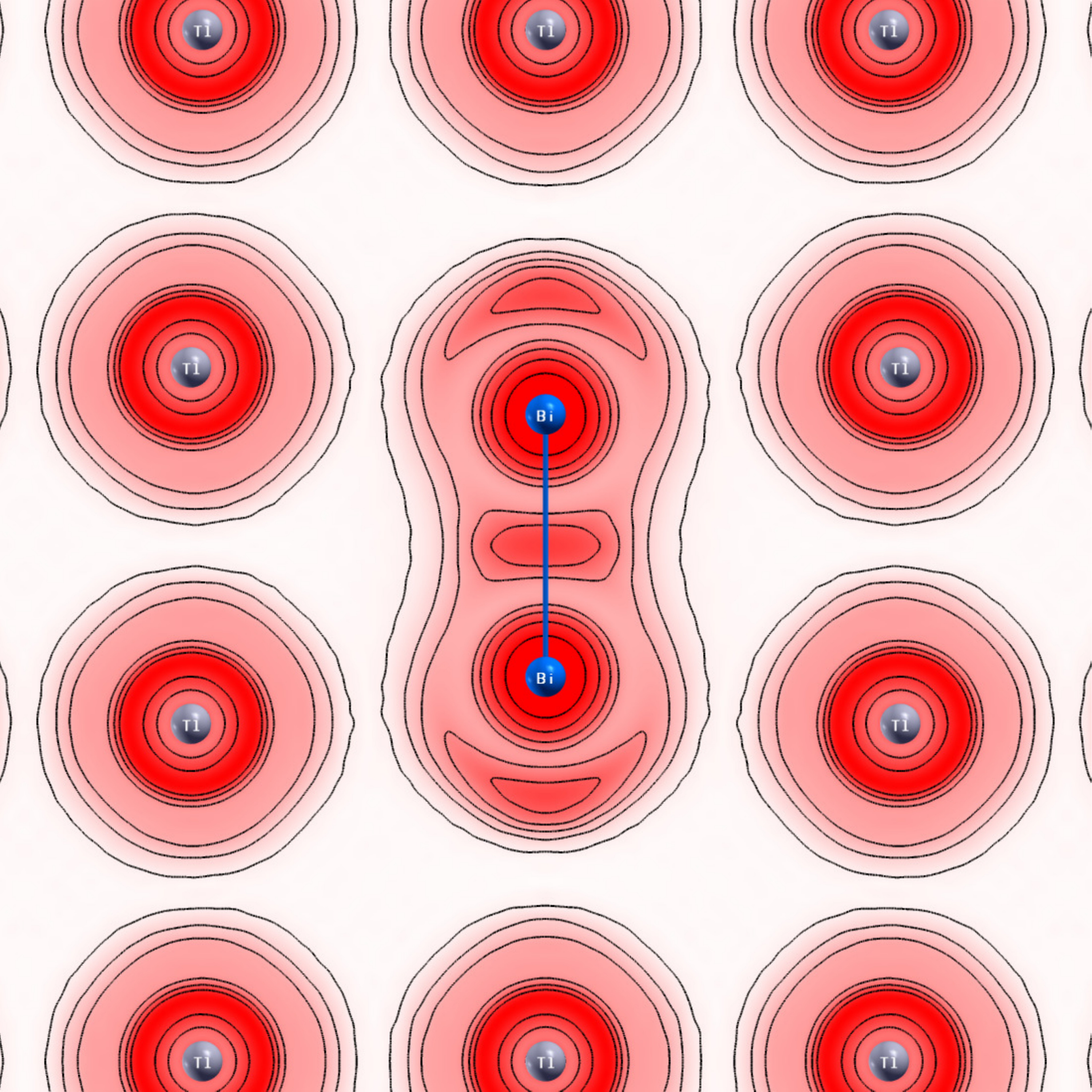}
\label{fig:Bi2+_TlCl_ELF}
}
\subfigure[]{%
\includegraphics[width=8.8cm, bb=0 0 1100 1100]{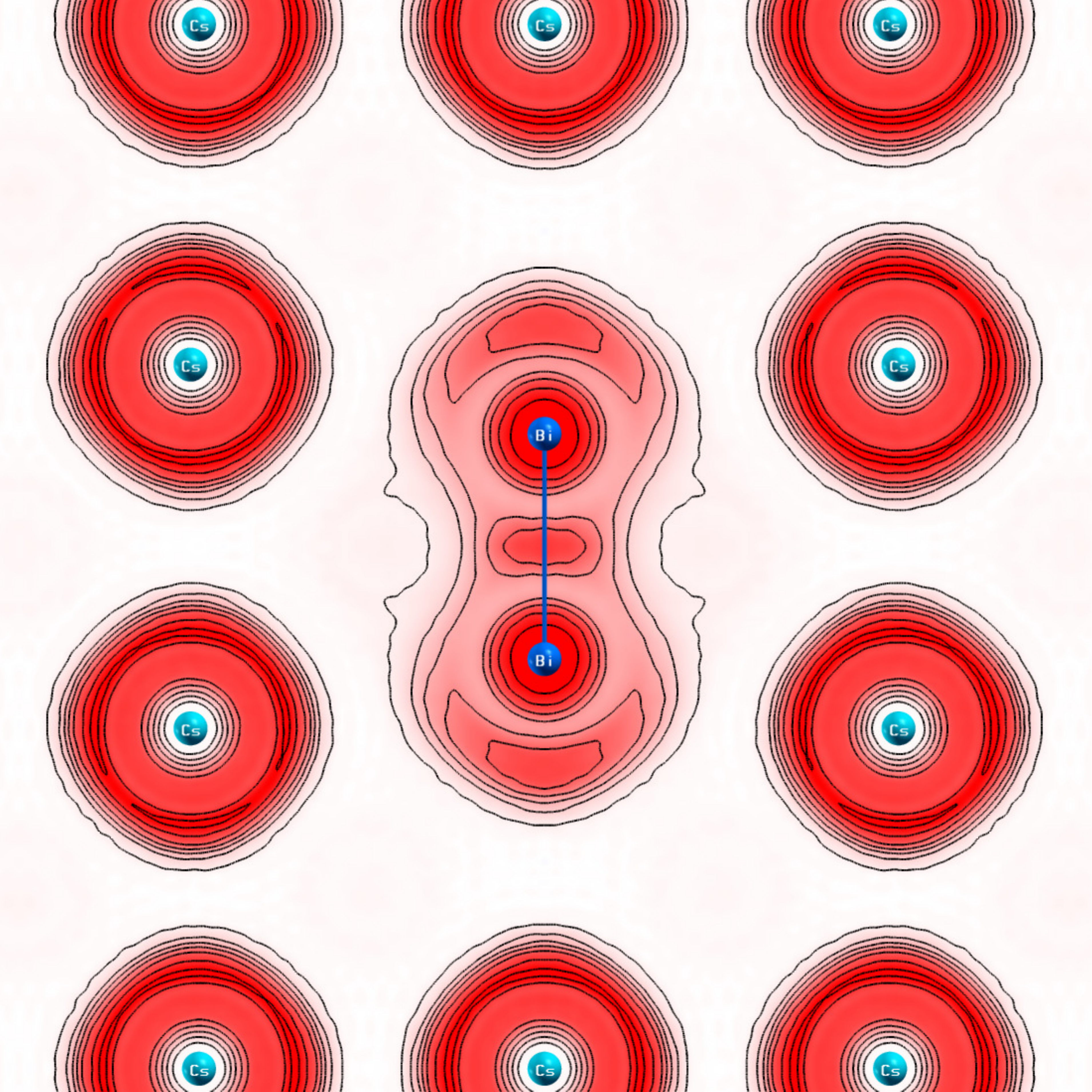}
\label{fig:Bi2+_CsI_ELF}
}
\\[-0.25\baselineskip]
\subfigure[]{%
\includegraphics[width=8.8cm, bb=0 0 1100 1100]{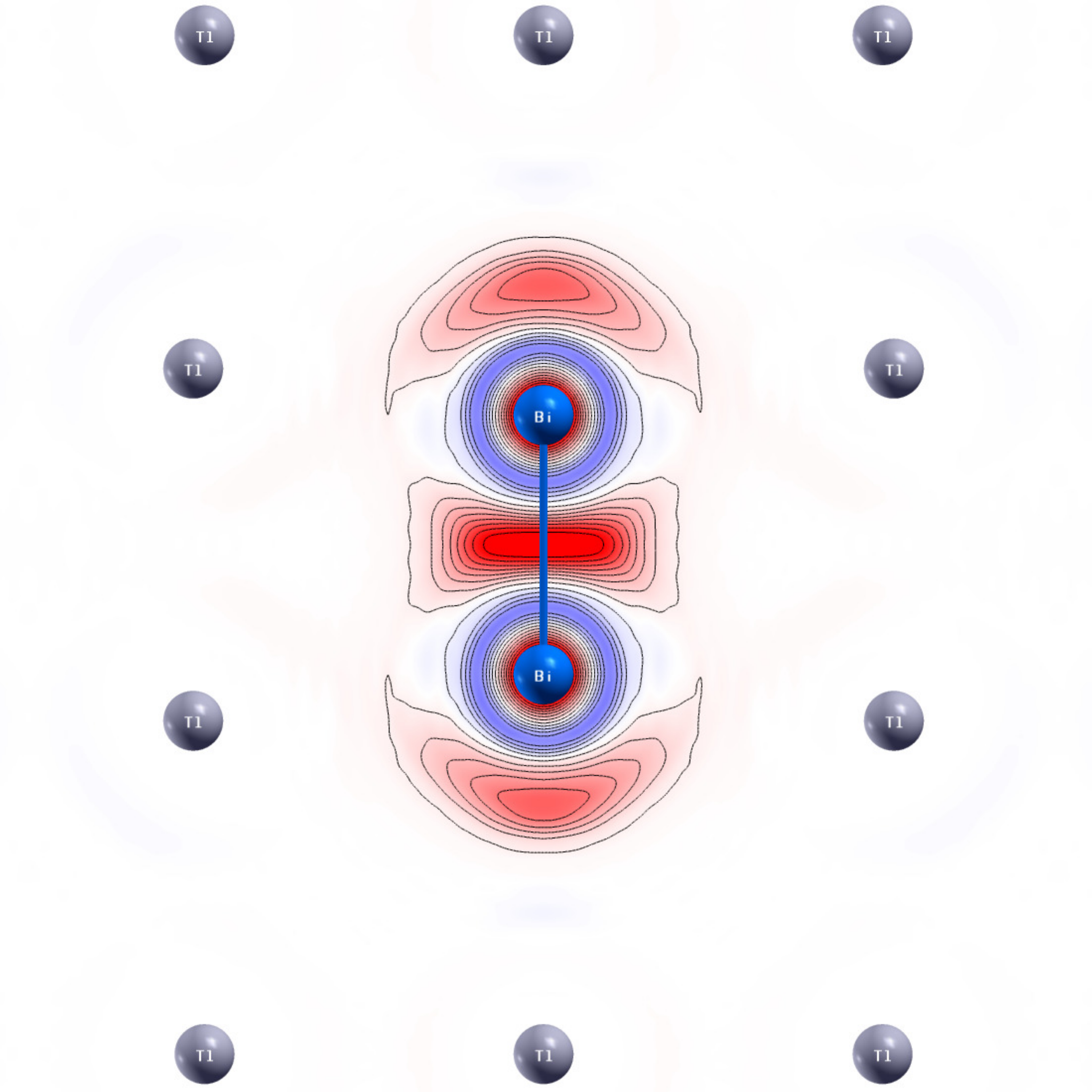}
\label{fig:Bi2+_minus_TlCl_ELF}
}
\subfigure[]{%
\includegraphics[width=8.8cm, bb=0 0 1100 1100]{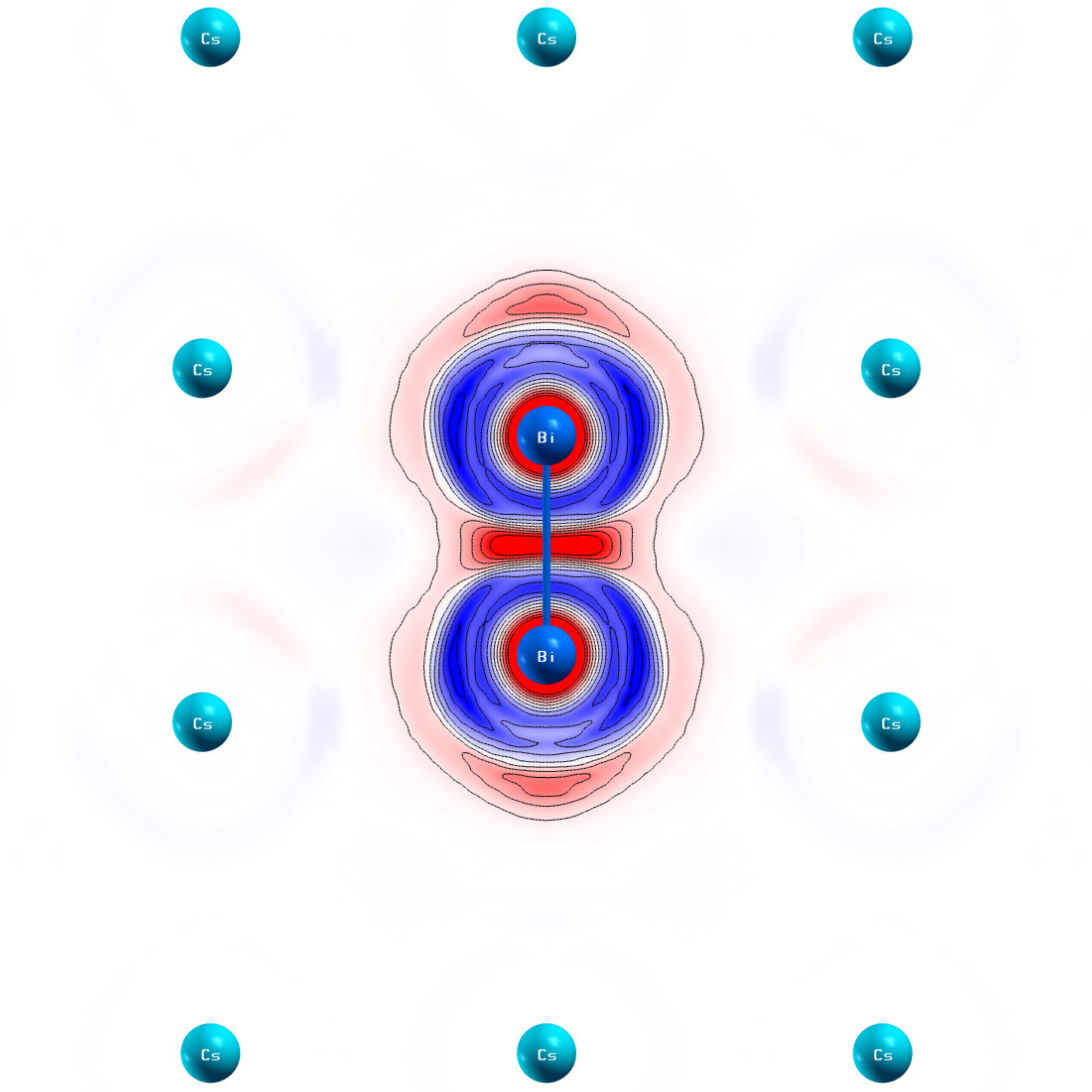}
\label{fig:Bi2+_minus_CsI_ELF}
}
\caption{%
Calculated electron localization functions of \Biiip{} dimer centers
in \subref{fig:Bi2+_TlCl_ELF}~\TlCl{} and \subref{fig:Bi2+_CsI_ELF}~\CsI, and
difference of calculated electron localization functions of \Biiip{} dimer
center from those of perfect lattice: \subref{fig:Bi2+_minus_TlCl_ELF}~\TlClBi,
\subref{fig:Bi2+_minus_CsI_ELF}~\CsIBi{} (in the $\left(101\right)$ plane).
}
\label{fig:Bi2+_ELFs}
\end{figure*}

\section{Results and discussion}
\label{sec:Results}
\subsection{Bi$^+$ substitutional centers}
\label{sec:Bi^+}
Calculation of \Bipi{} substitutional center shows that both in \TlCl{} and
in \CsI{} the crystal lattice is distorted rather slightly around it: bismuth
atom lies in the cation site, the nearest chlorine or iodine atoms are displaced
by almost $0.1a$ towards the bismuth cite, $a$ being the crystal lattice
constant ($a \approx 0.3834$~nm in \TlCl{} and $a \approx 0.4565$~nm in \CsI),
and the nearest thallium or cesium atoms are displaced apart from the bismuth
cite. So Bi$\relbar$Cl and Bi$\relbar$Tl distances are 0.3120 and 0.3889~nm,
respectively (0.3320 and 0.3834~nm in perfect \TlCl{} crystal) and Bi$\relbar$I
and Bi$\relbar$Cs distances are 0.3641 and 0.4624~nm, respectively (0.3955 and
0.4567~nm in perfect \CsI{} crystal).

Bader analysis of the electron density around \Bipi{} center in \TlCl{} showed
the atomic effective charges to be $+1.03\left|\textrm{e}\right|$,
$-0.70\left|\textrm{e}\right|$, and $+0.73\left|\textrm{e}\right|$ in -bismuth
atom, in each of the nearest chlorine atoms, and in each of the nearest
thallium atoms, respectively. The same analysis in \CsI{} yielded
$+0.75\left|\textrm{e}\right|$, $-0.67\left|\textrm{e}\right|$, and
$+0.80\left|\textrm{e}\right|$ atomic effective charges in bismuth atom, in each
of the nearest iodine atoms, and in each of the nearest cesium atoms,
respectively. The effective atomic charges in perfect \TlCl{} and \CsI{} crystal
lattices calculated by the same approach were found to be
$-0.73\left|\textrm{e}\right|$ and $+0.73\left|\textrm{e}\right|$ in chlorine
and thallium atoms, respectively, and $-0.80\left|\textrm{e}\right|$ and
$+0.80\left|\textrm{e}\right|$ in iodine and cesium atoms, respectively.

In \TlCl{} the total charge localized in cation cite and the neighboring anion
sites was changed from $-3.65\left|\textrm{e}\right|$ in perfect lattice to
$-3.17\left|\textrm{e}\right|$ in the lattice with bismuth substitutional
center. In CsI{} this total charge was changed from
$-4.00\left|\textrm{e}\right|$ in perfect lattice to
$-3.27\left|\textrm{e}\right|$ in the lattice with bismuth substitutional
center. This shows that the electron density was displaced into the space
between bismuth atom and the nearest anions and may be considered as slight
covalent contribution to Bi$\relbar$Cl and Bi$\relbar$I interaction. In \CsI{}
such displacement is more noticeable (Fig.~\ref{fig:Bi+_ELFs}).

The calculated energy levels of the \Bipi{} center in \TlClBi{} and \CsIBi{} are
shown in Figs.~\ref{fig:Bi+_TlCl_levels} and \ref{fig:Bi+_CsI_levels},
respectively, together with the corresponding transitions. It should be noticed
that splitting of the \Bipi{} ion states in bismuth substitutional centers in
\TlClBi{} and \CsIBi{} is not described by crystal field theory due to total
cubic symmetry of bismuth ion environment, unlike to the case of
\mbox{$\textrm{Tl}^0\!\left(1\right)$} center \cite{Mollenauer83}. In this case
the splitting is caused by electron density redistribution with a covalent
contribution formed. Absorption near 1.0, 0.8, 0.7, and $\sim 0.5$~\mkm{} is
found in the \Bipi{} center in \TlCl. The above-mentioned excited states
evaluative calculation allows one to expect a luminescence in \TlClBi{} near
1.0~\mkm{} excited in this absorption bands. In \CsIBi{} the IR luminescence in
the 1.2--1.3~\mkm{} range may be expected to be excited in absorption bands
near 1.2, 0.7, 0.5, and $\gtrsim 0.45$~\mkm. Besides, in \TlClBi{} another one
luminescence transition, with much a lower lifetime, may occur near 0.8~\mkm.

\subsection{Bi$_2^+$ dimer centers}
\label{sec:Bi_2^+}
The modeling shows that the \Biiip{} dimer centers can actually occur  both in
\CsI{} and \TlCl{} crystals.

In \Biiip{} centers the bismuth atoms are found to be displaced from the
adjacent cation sites by $0.12a$ and $0.18a$ towards each other, so that the
Bi$\relbar$Bi distance is reduced to 0.2903 and 0.2961~nm in \TlClBi{} and
\CsIBi, respectively. The nearest chlorine or iodine atoms are displaced towards
the dimer, and the nearest thallium or cesium atoms are displaced apart from the
dimer. As a result, the Bi$\relbar$Cl and Bi$\relbar$Tl distances are 0.3060 and
0.3960~nm, respectively (0.3320 and 0.3834~nm, respectively, in perfect \TlCl{}
crystal), and the Bi$\relbar$I, and Bi$\relbar$Cs distances become -0.3314 and
0.4743~nm, respectively, as compared to 0.3955 and 0.4567~nm,
respectively, in perfect \CsI{} lattice.

Bader's method analysis of electron density around the center shows that the
excess electron charge, $-1\left|\mathrm{e}\right|$, is localized almost
completely in the first coordination shell of two bismuth sites, mainly in
bismuth atoms and partially in the nearest chlorine or iodine atoms
(Fig.~\ref{fig:Bi2+_ELFs}). In \TlClBi{} the effective charge in each of the
bismuth atoms is found to be $+0.82\left|\textrm{e}\right|$, and the effective
charge in each of the nearest chlorine and thallium atoms is
$-0.73\left|\textrm{e}\right|$ and $+0.72\left|\textrm{e}\right|$, respectively.
So, as compared to \Bipi{} center, positive charge in each of the bismuth atoms
is decreased by $0.21\left|\textrm{e}\right|$, negative charge in each of
the neighboring chlorine atoms is increased by $0.03\left|\textrm{e}\right|$,
and positive charge in each of the nearest thallium atoms is decreased by
$0.01\left|\textrm{e}\right|$. Hence the excess total charge localized in two
bismuth atoms and their nearest neighbors turns out to be $\approx
-0.6\left|\textrm{e}\right|$. In \CsIBi{} the effective charge in each of the
bismuth atoms is $+0.48\left|\textrm{e}\right|$, and the effective charge in
each of the nearest iodine and cesium atoms is $-0.70\left|\textrm{e}\right|$
and $+0.79\left|\textrm{e}\right|$, respectively. As compared to \Bipi{} center,
positive charge in each of the bismuth atoms is decreased by
$0.27\left|\textrm{e}\right|$, negative charge in each of the neighboring iodine
atoms is increased by $0.03\left|\textrm{e}\right|$, and positive charge in each
of the nearest cesium atoms is decreased by $0.01\left|\textrm{e}\right|$. So
the excess total charge in two bismuth atoms and their environment is $\approx
-0.7\left|\textrm{e}\right|$. So the center is actually \Biiip{} dimer both in
\TlClBi{} and in \CsIBi. However, again a certain part of the electron density
is displaced into the space between bismuth atoms and the nearest anions, as in
\Bipi{} centers.

The calculated energy levels of \Biiip{} center in \TlCl{} and \CsI{} and the
corresponding transitions are shown in Figs.~\ref{fig:Bi_2+_TlCl_levels} and
\ref{fig:Bi_2+_CsI_levels}, respectively. Basing on the evaluative calculations
of the first excited state one might expect the IR luminescence bands in
\TlClBi{} near 1.1 and 1.3~\mkm{} excited in absorption near 1.1, 0.8, 0.7, 0.5
and $\sim 0.4$~\mkm, and in \CsIBi{} near 1.2 and 1.6~\mkm{} excited in
absorption near 1.2, 0.7, 0.6, and $\lesssim 0.5$~\mkm. One more luminescence
band near $\sim 2.4$~\mkm{} in \TlClBi{} and in the $\gtrsim 2.2$~\mkm{} range
in \CsIBi{} corresponding to a transition from the lowest excited state might be
observable at a low temperature.
\begin{figure*}
\subfigure[]{%
\includegraphics[width=8.8cm, bb=0 0 1100 1100]
{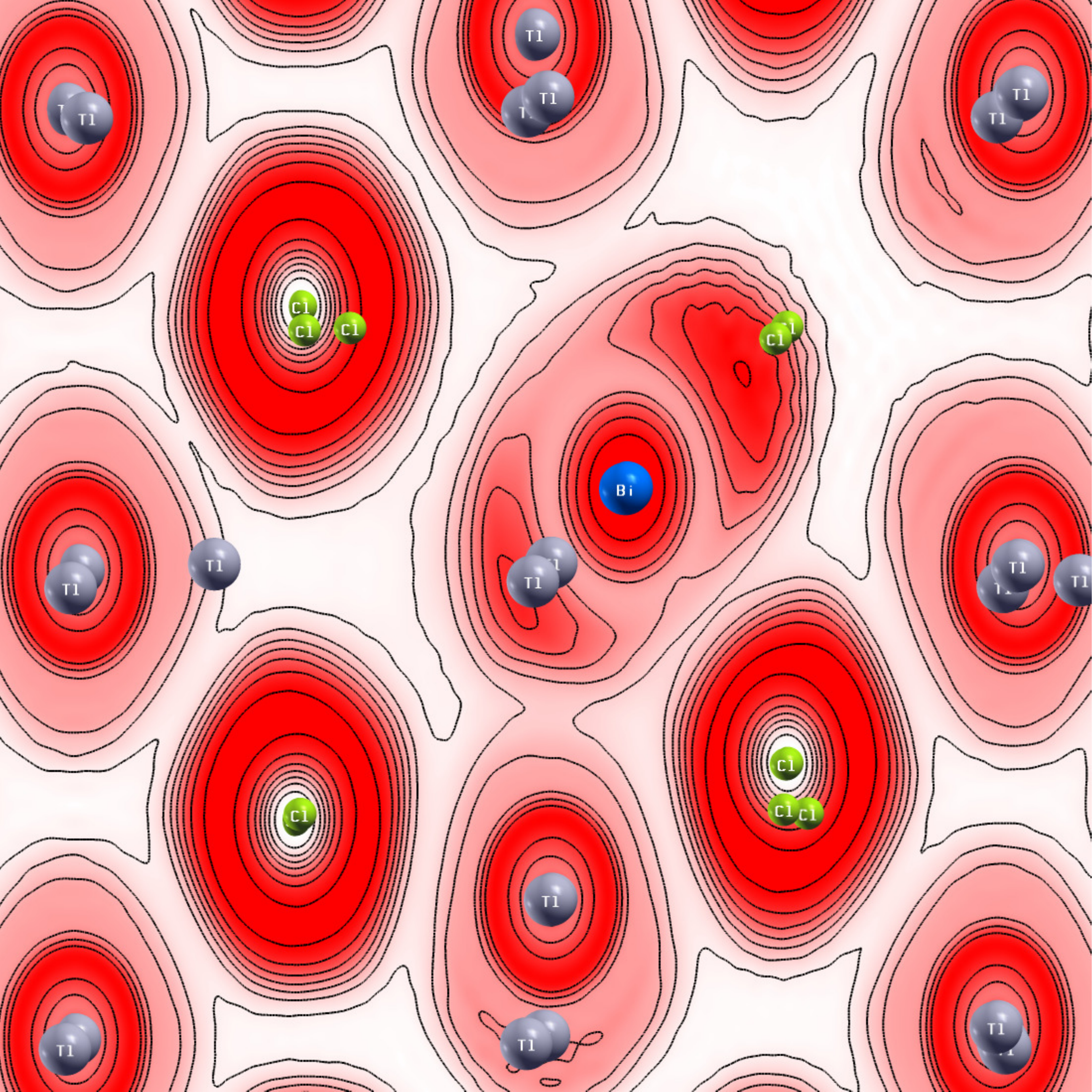}
\label{fig:Bi_and_vac_TlCl_ELF}
}
\subfigure[]{%
\includegraphics[width=8.8cm, bb=0 0 1100 1100]
{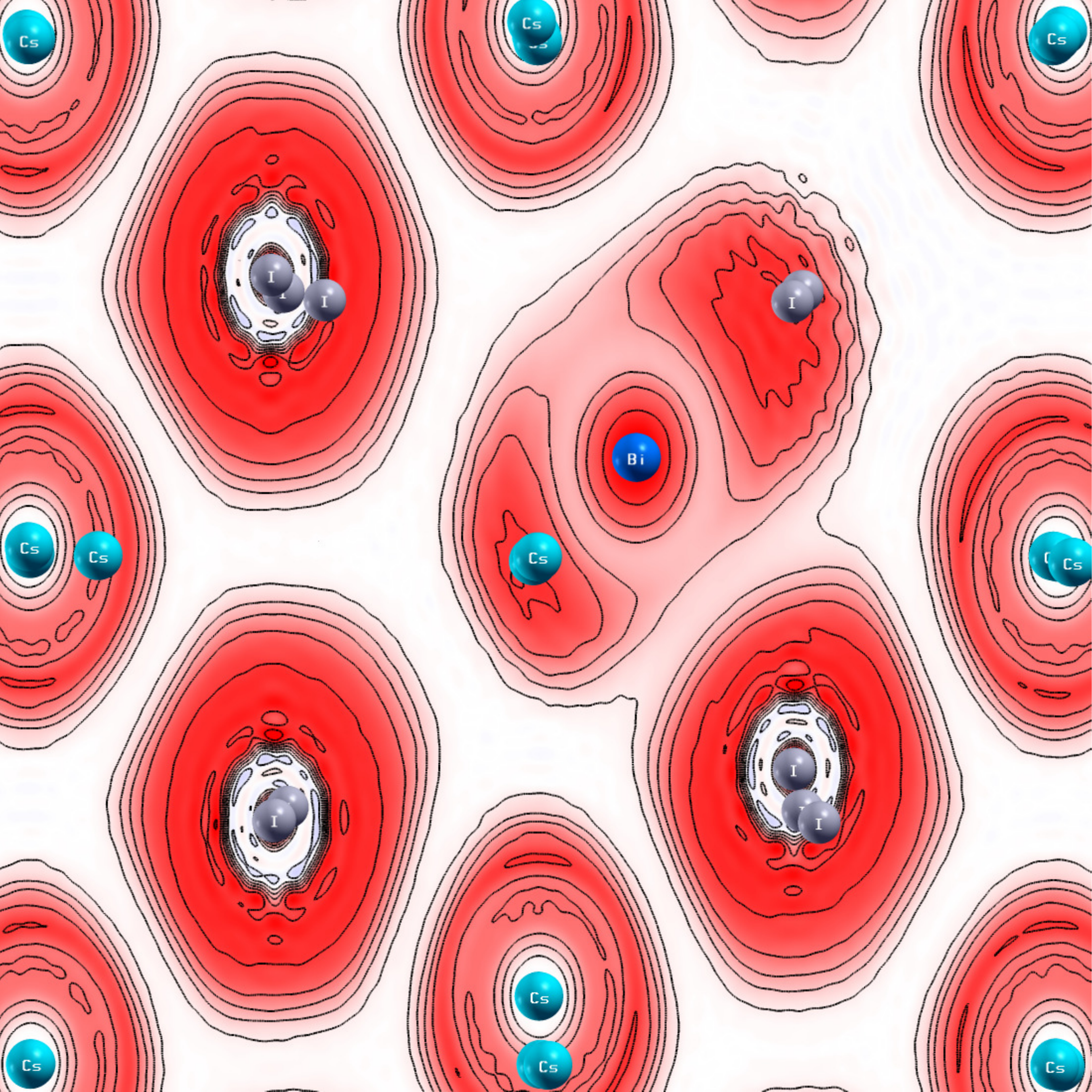}
\label{fig:Bi_and_vac_CsI_ELF}
}
\\[-0.25\baselineskip]
\subfigure[]{%
\includegraphics[width=8.8cm, bb=0 0 1100 1100]
{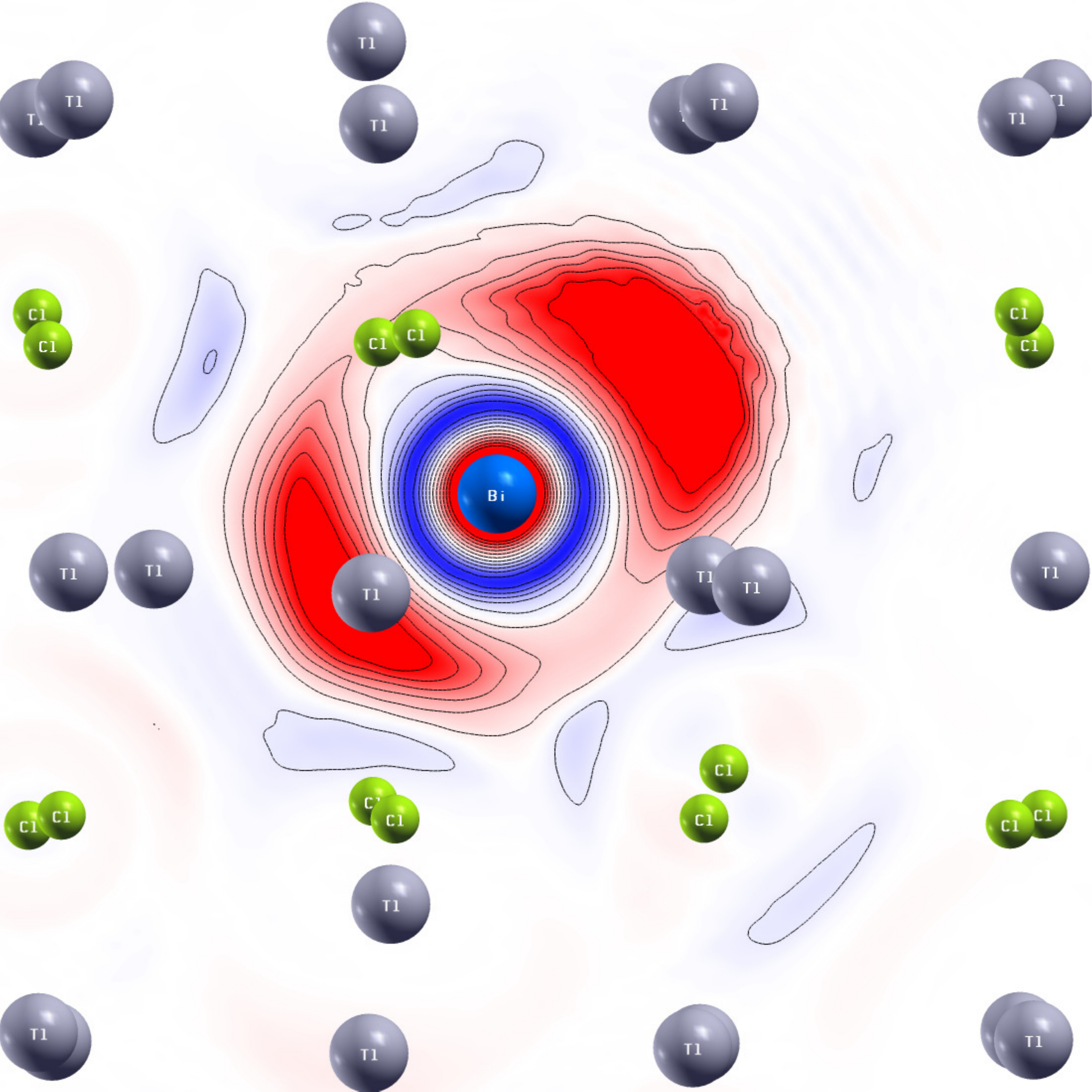}
\label{fig:Bi_and_vac_minus_TlCl_ELF}
}
\subfigure[]{%
\includegraphics[width=8.8cm, bb=0 0 1100 1100]
{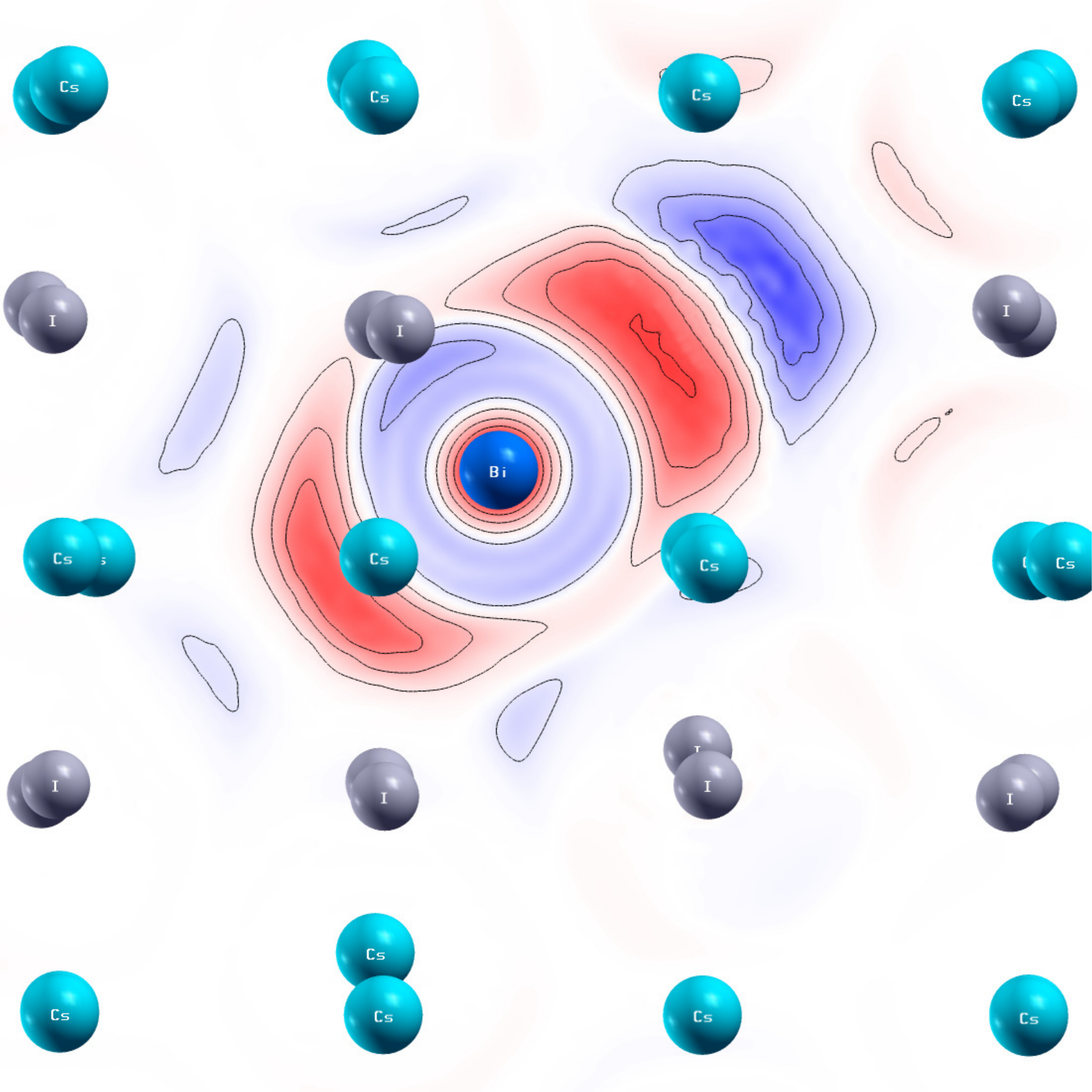}
\label{fig:Bi_and_vac_minus_CsI_ELF}
}
\caption{%
Calculated electron localization functions of
\subref{fig:Bi_and_vac_TlCl_ELF}~\BiVacCl{} center in \TlClBi{} and 
\subref{fig:Bi_and_vac_CsI_ELF}~\BiVacI{} center in \CsIBi{}  (in the
$\left(100\right)$ plane) and difference of calculated electron localization
functions of \subref{fig:Bi_and_vac_TlCl_ELF}~\BiVacCl{} center in \TlClBi{}
from those of perfect crystal and \subref{fig:Bi_and_vac_CsI_ELF}~\BiVacI{}
center in \CsIBi{} from those of perfect crystal (in the $\left(111\right)$
plane).
}
\label{fig:Bi_and_vac_ELFs}
\end{figure*}

\subsection{Bi substitutional -- anion vacancy complex centers}
\label{sec:Bi_V}
The modeling of the \BiVacz{anion}{} complexes shows that the complexes in
\TlCl{} and \CsI{} differ significantly in electronic and spectral properties.
The \BiVacz{Cl} complex in \TlClBi{} turns out to be similar in
certain respect to \mbox{$\textrm{Tl}^0\!\left(1\right)$} centers in alkali
halide crystals. However, the \BiVacz{I}{} complex in \CsI{} differs strikingly
from both the \BiVacz{Cl} and \mbox{$\textrm{Tl}^0\!\left(1\right)$} centers.

In the \BiVacz{Cl}{} complex in \TlCl{} the lattice relaxation turns out to be
significant. The bismuth atom is displaced by $0.30a$ from cation site towards
the vacant chlorine site, the nearest chlorine and thallium atoms are displaced
towards bismuth atom, and thallium atoms surrounding chlorine vacancy are
displaced apart from the vacant site. This results in  Bi$\relbar$Cl,
Bi$\relbar$Tl, and $\textrm{Tl}\!\relbar\!\textrm{V}_{\textrm{Cl}}$ distances
being 0.3024, 0.3598, and 0.4009~nm, respectively, as compared to the distances
0.3320, 0.3834, and 0.3320~nm in perfect \TlCl{} crystal. The relaxation is
accompanied by the electron density shifted from bismuth atom into the chlorine
vacancy region, so that the complex center may be thought of as a bound pair of
ions, {\lq\lq}\Bipi{} plus negatively charged V$_{\textrm{Cl}}^{-}$
vacancy{\rq\rq} (Fig.~\ref{fig:Bi_and_vac_TlCl_ELF}). Thus this center is a
\BiVacCl{} complex.

In \BiVacz{I}{} complex in \CsI, on the one hand, the lattice relaxation is as
significant as that in \BiVacCl{} center. Bismuth atom is displaced by $0.18a$
from the cation site towards the vacant iodine site, the nearest iodine atoms
are displaced towards bismuth atom, but the nearest cesium atoms are displaced
apart from bismuth atom. This results in Bi$\relbar$Cs and Bi$\relbar$I
distances decreased to 0.4220 and 0.3509~nm, respectively, from 0.4567 and
0.3955~nm, respectively, in perfect \CsI{} crystal. On the other hand, the
electron density distribution turns out to differ strikingly from that in
\BiVacCl{} complex in \TlCl{} crystal. As distinct from \TlCl, the electron
density in \CsI{} is not redistributed between bismuth atom and anion vacancy.
So the \BiVacz{I}{} complex may be roughly thought of as a bound pair: a neutral
substitutional atom, \Biz, and a neutral iodine vacancy, V$_{\textrm{I}}^{0}$
(Fig.~\ref{fig:Bi_and_vac_CsI_ELF}). In other words, this center turns out to be
a \BiVacI{} complex.

Bader's method analysis of the electron density distribution around the center
confirms these conclusions. In \TlCl{} the effective charge of bismuth atom is
$+0.79\left|\textrm{e}\right|$, effective charge of each of the nearest chlorine
atoms is $-0.70\left|\textrm{e}\right|$, and effective charge of each of the
thallium atoms surrounding the chlorine vacancy is
$+0.66\left|\textrm{e}\right|$. So the excess (in comparison with perfect
\TlCl{} lattice) positive charge localized in bismuth substitutional atom and
its nearest neighbors is about $0.85\left|\textrm{e}\right|$, and the negative
charge no less than $0.4\left|\textrm{e}\right|$ is localized in the chlorine
vacancy and in its nearest neighbors. In \CsI{} the effective charge of bismuth
atom is $-0.02\left|\textrm{e}\right|$, effective charge of each of the nearest
iodine atoms is $-0.82\left|\textrm{e}\right|$, and effective charge of each of
the cesium atoms surrounding the iodine vacancy is
$+0.79\left|\textrm{e}\right|$. So the bismuth substitutional atom turns out to
be practically a neutral atom, the excess (in comparison with perfect \CsI{}
lattice) positive charge localized in its nearest neighbors is about
$0.08\left|\textrm{e}\right|$ and the positive charge no less than
$0.06\left|\textrm{e}\right|$ is localized in the iodine vacancy and in its
nearest neighbors.

The calculated energy levels of \BiVacCl{} and \BiVacI{} centers and the
corresponding transitions are shown in Figs.~\ref{fig:Bi_and_vac_TlCl_levels}
and \ref{fig:Bi_and_vac_CsI_levels}, respectively. Spectral properties of the
\BiVacCl{} and \BiVacI{} complexes may be understood in a model similar to
\mbox{$\textrm{Tl}^0\!\left(1\right)$} center theory \cite{Mollenauer83}. In
such a model the complexes are considered as \Bipi{} ion or \Biz{} atom,
respectively, in the axial crystal field formed by neighboring chlorine or
iodine vacancy. Obviously, the crystal field of a negatively changed chlorine
vacancy is stronger than that of a neutral iodine vacancy.

Three lowest states of \Bipi{} ion are known to arise from \Term{3}{P}{}{}
atomic state split by strong spin-orbital interaction in bismuth. The ground
state of \Bipi{} ion is \Term{3}{P}{0}{}. The first excited state,
\Term{3}{P}{1}{}, is split by an axial crystal field in two levels, 6100 and
8300~\cminv, and the second excited state, \Term{3}{P}{2}{}, is split in three
levels, 12500, 14900 and 25000~\cminv. In a free \Bipi{} ion the electric dipole
(E1) transitions between three spin-orbital components of the \Term{3}{P}{}{}
state are forbidden but under the influence of the crystal field the transitions
become allowed. Basing on the above-mentioned evaluative calculations one
expects the IR luminescence in two bands near 1.6 and 1.2~\mkm, both excited in
absorption near 0.8, 0.7, and $\sim 0.4$~\mkm{}. A luminescence band with a
significantly shorter (by an order of magnitude) lifetime may occur near
0.8~\mkm.

The ground state of \Biz{} atom is known to be \Term{4}{S}{3/2}{}. The first
excited state, \Term{2}{D}{3/2}{}, is split by an axial crystal field in two
levels, 9000 and 10100~\cminv. The second excited state, \Term{2}{D}{5/2}{}, is
split in three levels, 16000, 18000, and 19300~\cminv, and the forth excited
state, \Term{2}{P}{1/2}{}, is not split by an electrostatic field. Notice that
the splitting is expected to be considerably slighter than that in \BiVacCl{}
complex. E1 transitions from the ground state of \BiVacI{} center, corresponding
to the \Term{4}{S}{3/2}{} atomic state, to all the states arising from the
\Term{2}{D}{}{} one, turn out to be weak since in a free atom such transitions
are parity-forbidden. Hence the only relatively intensive absorption band near
0.43~\mkm{} corresponding to
\Term{4}{S}{3/2}{}$\,\rightarrow\,$\Term{2}{P}{1/2}{} transition may be expected
to occur in the \BiVacI{} center (Fig.~\ref{fig:Bi_and_vac_CsI_levels}). With
the above-mentioned evaluative calculations taken into account one might expect
that 1.0 -- 1.1~\mkm{} luminescence corresponding to
\Term{2}{D}{3/2}{}$\,\rightarrow\,$\Term{4}{S}{3/2}{} transitions is excited in
this absorption band.

\section{Conclusions}
\label{sec:Conclusions}
Spectroscopic data \cite{We13, Su11, Su12} and the present results of modeling
of bismuth-related centers in \TlClBi{} and \CsIBi{} crystals suggest that
different centers are responsible for the near-infrared luminescence in these
crystals.

In \TlClBi{} the luminescence observed near 1.2~\mkm{} is caused mainly by
\BiVacz{Cl}{} complexes formed by \Bipi{} substitutional ions and negatively
charged chlorine vacancies. However, the bismuth-related contribution to the
total absorption is caused mainly by single \Bipi{} substitutional centers not
contributing to the IR luminescence. This explains a significant difference
between the IR luminescence excitation spectrum and \TlClBi{} absorption spectra
\cite{We13}. \Bipi{} substitutional centers may cause a luminescence near
1.0~\mkm{} not observed in \cite{We13}.

In \CsIBi{} the IR luminescence is caused mainly by two types of bismuth
centers: \Bipi{} substitutional centers give rise to a luminescence observed
in the 1.2 -- 1.3~\mkm{} range, and \Biiip{} dimer centers cause a luminescence
observed near 1.2~\mkm{} and near 1.6~\mkm. These conclusions agree with the
experimental data of \cite{Su11, Su12} and confirm the assumptions made there.
Besides, dimer centers may cause possible luminescence in the $\gtrsim
2.2$~\mkm{} range not observed in Refs.~\cite{Su11, Su12}.

\Biiip{} dimer complexes in \TlClBi{} can contribute perceptibly neither to the
IR luminescence spectra nor to the absorption spectra, as distinct from \CsIBi{}
crystals.

\BiVacz{I}{} complexes formed in \CsIBi{} by \Biz{} substitutional atoms and
neutral iodine vacancies may contribute to the IR luminescence only near
1.0~\mkm, as distinct from \TlClBi{} crystals. One might suppose that the
luminescence in the 0.95 -- 1.15~\mkm{} range observed in \cite{Su11} at a low
temperature under 0.4 -- 0.5~\mkm{} excitation is contributed by \BiVacI{}
centers formed due to electrons capturing in \Bipi{} substitutional centers in
the vicinity of iodine vacancies.

\acknowledgments
This work is supported in part by Fundamental Research Program of the Presidium
of the Russian Academy of Sciences and by Russian Foundation for Basic Research
(grant \mbox{12-02-00907}).

%
%

\end{document}